\documentclass[leqno,draft,11pt]{article}
\usepackage{amssymb,amsmath,latexsym,theorem,a4wide}
\usepackage[final]{graphicx}

\allowdisplaybreaks[2]

\let\mr\mathrel

\let\eq\leftrightarrow
\def\iff{\quad\text{iff}\quad}
\let\ET\bigwedge
\let\TO\Rightarrow
\let\model\vDash
\let\nmodel\nvDash

\let\fii\varphi

\def\p#1{\langle#1\rangle}
\def\lh#1{\lvert#1\rvert}
\let\sset\subseteq
\let\Sset\supseteq
\def\res{\mathbin\restriction}

\let\dia\Diamond
\def\ru{\mathrel/}

\DeclareRobustCommand*\adm{%
  \mathrel{%
   \setbox0 \hbox{$\mathop\vdash$}\dimen0 \ht0
   \setbox0 \hbox{$\vdash$}\advance\dimen0 -\ht0
   \vrule width.8\fontdimen8 \textfont3 height\ht0 depth\dp0
   \mkern-1mu
   \lower\dimen0 \hbox{$\vcenter{\hbox{$\scriptstyle\sim\mathstrut$}}$}}}

\DeclareMathOperator\md{md}

\let\lgc\mathbf
\def\alc{\mathcal{ALC}}

\def\bme{\hskip.75em\relax}
\def\noproof{\leavevmode\unskip\bme\vadjust{}\nobreak\hfill$\qed$\par}
\let\qed\Box
\newenvironment{Pf}
  {\par\noindent\textit{Proof:}\bme\ignorespaces}
  {\noproof\pagebreak[2]\vskip\medskipamount\ignorespacesafterend}

\theoremstyle{plain}
\newtheorem{Thm}{Theorem}[section]
\newtheorem{Prop}[Thm]{Proposition}
\newtheorem{Cor}[Thm]{Corollary}
\newtheorem{Lem}[Thm]{Lemma}
\newtheorem{Fact}[Thm]{Fact}
\theorembodyfont\upshape
\newtheorem{Def}[Thm]{Definition}

\author{Emil Je\v r\'abek\\[\medskipamount]
Institute of Mathematics of the Academy of Sciences\\
\small \v Zitn\'a 25,
115\:67 Praha 1,
Czech Republic,
email: \texttt{jerabek@math.cas.cz}
}

\title{Blending margins: The modal logic $\lgc K$ has nullary unification type}

\begin{document}
\maketitle
\begin{abstract}
We investigate properties of the formula $p\to\Box p$ in the basic
modal logic~$\lgc K$. We show that $\lgc K$ satisfies an infinitary
weaker variant of the rule of margins
$\fii\to\Box\fii\ru\fii,\neg\fii$, and as a consequence, we obtain various
negative results about admissibility and unification in~$\lgc K$. We describe
a complete set of unifiers (i.e., substitutions making the formula
provable) of~$p\to\Box p$, and use it to establish that 
$\lgc K$ has the worst possible unification type: nullary.
In well-behaved transitive modal logics, admissibility and unification
can be analyzed in terms of projective formulas, introduced by
Ghilardi; in particular, projective formulas coincide for these logics with formulas
that are admissibly saturated (i.e., derive all their
multiple-conclusion admissible consequences) or exact (i.e.,
axiomatize a theory of a substitution). In contrast,
we show that in~$\lgc K$, the formula~$p\to\Box p$ is admissibly
saturated, but neither projective nor exact. All our results for~$\lgc
K$ also apply to the basic description logic~$\alc$.

\end{abstract}

\section{Introduction}

Equational unification studies the problem of making terms equivalent
modulo an equational theory by means of a substitution. It has
been thoroughly investigated for basic algebraic theories, such as the
theory of commutative semigroups, see Baader and Snyder~\cite{baa-sny}
for an overview. If $L$ is a propositional logic algebraizable with
respect to a class of algebras~$V$, unification modulo the equational
theory of~$V$ can be stated purely in terms of propositional logic: an
$L$-unifier of a set of formulas~$\Gamma$ is a substitution which
turns all formulas from~$\Gamma$ into $L$-tautologies.

In the realm of modal logics, the seminal results of
Ghilardi~\cite{ghil} show that unification is at most finitary,
decidable, and generally well-behaved for a representative class of
transitive modal logics, including e.g.\ $\lgc{K4}$, $\lgc{S4}$,
$\lgc{GL}$,~$\lgc{Grz}$. Unification was also studied for fragments of
description logics, which have applications in ontology generation and maintenance; see
Baader and Ghilardi~\cite{baa-ghi}. In particular, the description logics treated in \cite{baa-nar,baa-mor} 
can be thought of as the $\{\land,\Box\}$ and $\{\land,\dia\}$ fragments of multimodal~$\lgc K$.

Unification in propositional logics is closely connected to
admissibility of inference rules: a multiple-conclusion
rule~$\Gamma\ru\Delta$ is $L$-admissible if every $L$-unifier
of~$\Gamma$ also unifies some formula from~$\Delta$.
Rybakov~\cite{ryb:bk} proved that admissibility is decidable for a
class of transitive modal logics (similar to the one mentioned above)
and provided characterizations of their admissible rules. Some of
these results can be alternatively obtained using Ghilardi's approach
(cf.\ also~\cite{ejadm}). It is also possible to treat intuitionistic
and intermediate logics in parallel with the transitive modal
case~\cite{ryb:bk,ghilil,iem:aripc}.

In contrast to these results, not much is known about unification and
admissibility in nontransitive modal logics with a complete set of Boolean connectives. In particular, one of the
main open problems in the area is decidability of unification or
admissibility in the basic modal logic~$\lgc K$. (Wolter and
Zakharyaschev~\cite{wolt-zakh:undec} have shown that unifiability is
undecidable in the bimodal extension of~$\lgc K$ with the universal
modality and in some description logics, but it is wide open whether
one can extend these results to~$\lgc K$ itself.)

In this note we present some negative properties of unification and
admissibility in~$\lgc K$. The main result is that unification
in~$\lgc K$ is nullary (i.e., of the worst possible type). In terms of
description logic, unification in~$\alc$ is nullary, even if we
consider formulas with only one role and one concept name.
We also
show that there exists a formula (namely, $p\to\Box p$) which is
admissibly saturated in the sense of~\cite{ej:lukbas}, but it is not
projective (or even exact). In contrast, the results of
Ghilardi~\cite{ghil} imply that in well-behaved transitive modal
logics such as~$\lgc{K4}$, projective, exact, and admissibly saturated
formulas coincide, and indeed this is an important precondition which
makes possible the characterization of admissibility in terms of
projective approximations. Thus, admissible rules of~$\lgc K$ cannot
be directly analyzed in a similar way.

Our results are based on a classification of unifiers of the formula
$p\to\Box p$. The main ingredient is establishing that $\lgc K$ admits
a weaker version of the so-called rule of margins
\[\fii\to\Box\fii\ru\fii,\neg\fii\]
(meaning that whenever a formula of the form $\fii\to\Box\fii$ is valid, one of the formulas $\fii$, $\neg\fii$ is also
valid).
The rule of margins was investigated by
Williamson~\cite{will-alt,will-adm,will-mac} in the context of
epistemic logic. (The rule is supposed to express the ubiquity of
vagueness. We read~$\Box$ as ``clearly''. Since all our learning
processes have a certain margin of error, the only way we can know for
sure that $\fii$ is clearly true whenever it is true is that we know
in fact whether $\fii$ is true or false.) The rule of margins is
admissible e.g.\ in the logics $\lgc{KD}$, $\lgc{KT}$, $\lgc{KDB}$,
and~$\lgc{KTB}$, but not in~$\lgc K$. However, we will show that $\lgc
K$ satisfies a variant of the rule whose conclusion is that either
$\fii$ holds, or it is almost contradictory in the sense of
implying~$\Box^n\bot$ for some~$n\in\omega$. We remark that the rule of margins was also used in connection with
unification by Dzik~\cite{dzik:selfconj}.

\section{Preliminaries}

We refer the reader to~\cite{cha-zax,brv,baa-sny} for background on modal
logic and unification.
We review below the needed definitions to fix the notation, and some
relevant basic facts.

We work with formulas in the propositional modal language using
propositional variables~$p_n$ for~$n<\omega$ (we will often write
just~$p$ for~$p_0$), Boolean connectives (including
the nullary connectives~$\bot,\top$), and the unary modal
connective~$\Box$. We will use lower-case Greek letters~$\fii,\psi,\dots$ to denote
formulas, and upper-case Greek letters~$\Gamma,\Delta,\dots$ for finite sets of formulas. We
define $\dia\fii$, $\Box^n\fii$, $\Box^{<n}\fii$, and~$\dia^n\fii$ as
shorthands for $\neg\Box\neg\fii$,
$\underbrace{\Box\cdots\Box}_{n\text{ boxes}}\fii$,
$\ET_{i=0}^{n-1}\Box^i\fii$, and~$\neg\Box^n\neg\fii$, respectively.
(As a special case, $\Box^0\fii=\fii$ and $\Box^{<0}\fii=\top$.) The
\emph{modal degree}~$\md(\fii)$ of a formula~$\fii$ is defined so that
$\md(p_i)=0$,
$\md(\circ(\fii_0,\dots,\fii_{k-1}))=\max_{i<k}\md(\fii_i)$ for a
$k$-ary Boolean connective~$\circ$, and $\md(\Box\fii)=1+\md(\fii)$.

We use~$\vdash$ to denote the \emph{global consequence relation}
of~$\lgc K$. That is, $\Gamma\vdash\fii$ iff there exists a sequence
of formulas $\fii_0,\dots,\fii_n$ such that~$\fii_n=\fii$, and
each~$\fii_i$ is an element of~$\Gamma$, a classical propositional
tautology, an instance of the axiom
\[\Box(\alpha\to\beta)\to(\Box\alpha\to\Box\beta),\]
or it is derived from some of the formulas~$\fii_j$ with~$j<i$ by an
instance of necessitation $\alpha\ru\Box\alpha$ or modus ponens
$\alpha,(\alpha\to\beta)\ru\beta$.

A \emph{Kripke model} is a triple~$\p{F,R,{\model}}$, where the
\emph{accessibility relation}~$R$ is a
binary relation on a set~$F$, and the \emph{valuation}~$\model$ is a
relation between elements of~$F$ and formulas, written as
$F,x\model\fii$, which commutes with propositional connectives and
satisfies
\[F,x\model\Box\fii\iff\forall y\in F\,(x\mr Ry\TO F,y\model\fii).\]
If there is no danger of confusion, we will denote the
model~$\p{F,R,\model}$ by just~$F$.
We write $F\model\fii$ if $F,x\model\fii$ for every~$x\in F$, and
$F\model\Gamma$ if $F\model\fii$ for every~$\fii\in\Gamma$. The
\emph{strong completeness theorem} for~$\lgc K$ \cite[Thms.~3.55, 10.5]{cha-zax} states
\begin{Fact}\label{fact:fsc}
$\Gamma\vdash\fii$ iff $F\model\Gamma$ implies $F\model\fii$
for every model~$\p{F,R,\model}$.
\end{Fact}
We write $R(x)=\{y:x\mr Ry\}$. Let
$$R^n=\{\p{x_0,x_n}\in F^2:
\exists x_1,\dots,x_{n-1}\in F\,\forall i<n\,x_i\mr Rx_{i+1}\}$$
be the $n$-fold composition of~$R$ (where the case~$n=0$ is
understood to mean $R^0=\{\p{x,x}:x\in F\}$), and $R^{\le
n}=\bigcup_{i\le n}R^i$. We say that~$x$ is a
\emph{root} of~$F$ if $F=\bigcup_{n\in\omega}R^n(x)$.
\begin{Fact}[{{\cite[Cor.~3.29]{cha-zax}, cf.\ \cite[Thm.~2.34]{brv}}}]\label{fact:tree}
If $\nvdash\fii$, then there exists a model~$\p{F,R,\model}$ based on
a finite irreflexive intransitive tree with root~$x$ such that
$F,x\nmodel\fii$.
\end{Fact}
(That is, $R$ is the edge relation of a directed tree with edges
oriented away from~$x$ and no self-loops.)

A model~$\p{F',R',\model'}$ is the restriction of~$\p{F,R,\model}$
to~$F'$, denoted as~$\p{F,R,\model}\res F'$, if $F'\sset F$, $R'=R\cap
F'^2$, and $F,x\model p_i$ iff $F',x\model p_i$ for every $x\in F'$
and~$p_i$.
\begin{Fact}[{{\cite[Prop.~3.2]{cha-zax}, \cite[L.~2.33]{brv}}}]\label{fact:nbr}
If $n\ge\md(\fii)$, $x\in F\cap G$, and
$\p{F,R,\model}\res R^{\le n}(x)=\p{G,S,\model}\res S^{\le n}(x)$,
then $F,x\model\fii$ iff $G,x\model\fii$.
\end{Fact}
A \emph{p-morphism} between models $\p{F,R,\model}$
and~$\p{G,S,\model}$ is a function $f\colon F\to G$ such that
\begin{enumerate}
\item $x\mr R y$ implies $f(x)\mr Sf(y)$,
\item if $f(x)\mr Sz$, there exists~$y\in F$ such that $x\mr R y$ and~$f(y)=z$,
\item $F,x\model p_i$ iff $G,f(x)\model p_i$ for every variable~$p_i$.
\end{enumerate}
\begin{Fact}[{{\cite[Thm.~3.15]{cha-zax}, \cite[Prop.~2.14]{brv}}}]\label{fact:pmor}
If $f\colon F\to G$ is a p-morphism, then $F,x\model\fii$ iff
$G,f(x)\model\fii$ for every formula~$\fii$.
\end{Fact}

A \emph{substitution} is a mapping from formulas to formulas which
commutes with all connectives. A \emph{unifier} of a finite set of
formulas~$\Gamma$ is a substitution~$\sigma$ such that $\vdash\sigma(\fii)$ for
all~$\fii\in\Gamma$. In logics with a well-behaved conjunction
connective such as~$\lgc K$, unifiers of~$\Gamma$ are the same as
unifiers of the single formula~$\ET\Gamma$, hence we will mostly
restrict the discussion below to plain formulas instead of sets in
order to simplify the notation.

Let $U(\fii)$ be the set of all unifiers
of~$\fii$. The \emph{composition} of substitutions~$\sigma,\tau$ is
the substitution $\sigma\circ\tau$ such that
$(\sigma\circ\tau)(\fii)=\sigma(\tau(\fii))$. Let $\sigma\equiv\tau$
if $\vdash\sigma(p_i)\eq\tau(p_i)$ for every~$i$. A
substitution~$\tau$ is \emph{more general} than~$\sigma$, written
as~$\sigma\preceq\tau$, if there exists a substitution~$\upsilon$ such
that~$\sigma\equiv\upsilon\circ\tau$. We warn the reader that $\preceq$ is often written in the opposite direction in
literature on unification theory.
We write~$\sigma\approx\tau$ if
$\sigma\preceq\tau$ and~$\tau\preceq\sigma$, and $\sigma\prec\tau$ if
$\sigma\preceq\tau$ but~$\tau\npreceq\sigma$. Note that $\preceq$ is a
preorder, and $\approx$ is the induced equivalence relation. A
\emph{complete set of unifiers of~$\fii$} is a cofinal subset~$C$
of~$\p{U(\fii),\preceq}$ (i.e., a set of unifiers of~$\fii$ such
that every unifier of~$\fii$ is less general than some element
of~$C$). If $\{\sigma\}$ is a complete set of unifiers of~$\fii$,
then $\sigma$ is a \emph{most general unifier (mgu)} of~$\fii$.

If $\p{P,\le}$ is a nonempty poset, let $M$ be the set of its maximal
elements (i.e., $x\in P$ such that $x<y$ for no~$y\in P$). If every
element of~$P$ is below an element of~$M$, we say that $\p{P,\le}$ is
of
\begin{itemize}
\item type~$1$ (unitary), if $\lh M=1$,
\item type~$\omega$ (finitary), if $M$ is finite and~$\lh M>1$,
\item type~$\infty$ (infinitary), if $M$ is infinite.
\end{itemize}
Otherwise, it is of type~$0$ (nullary).

The \emph{unification type} of~$\fii$ is the type of the quotient
poset~$\p{U(\fii),\preceq}/{\approx}$. Note that $\fii$ is of
unitary type iff it has an mgu, and it is of at most finitary type
(i.e., $1$ or~$\omega$) iff it has a finite complete set of unifiers.
The unification type of a logic (that
is, for us, of~$\lgc K$) is the maximal type of a unifiable
formula~$\fii$, where we order the unification types as
$1<\omega<\infty<0$.

In unification theory, it is more customary to define the equivalence of unifiers~$\sigma,\tau\in U(\fii)$ (and derived
notions such as $\preceq$ and unification types) so that $\sigma\equiv\tau$ iff $\vdash\sigma(p_i)\eq\tau(p_i)$ for
variables~$p_i$ \emph{that occur in~$\fii$,} whereas we demanded this for all variables. Our results hold equally well under
the restricted definition, and in fact, the proofs could be slightly simplified in this case (we could replace
conditions \ref{item:4}, \ref{item:5} in Lemma~\ref{lem:sigtop} with just $\sigma\equiv\sigma_\top$). The latter is one
reason for our choice of the definition: in order to make the results most general, we carry out the proofs for the
most complicated case. We also find it convenient to have an absolute notion of equivalence of substitutions,
independent of which formula they are considered to be unifiers of. Our results are robust under further variations
of the definition, for example we could consider substitutions with domain consisting of formulas using only variables
occurring in~$\fii$, and target consisting of formulas using variables from a fixed finite set (which could be the
same as the domain).

A \emph{multiple-conclusion rule} is an expression~$\Gamma\ru\Delta$,
where $\Gamma,\Delta$ are finite sets of formulas. A
rule~$\Gamma\ru\Delta$ is \emph{derivable} if $\Gamma\vdash\psi$ for
some~$\psi\in\Delta$. A rule~$\Gamma\ru\Delta$ is \emph{admissible},
written as~$\Gamma\adm\Delta$, if every unifier of~$\Gamma$ also
unifies some~$\psi\in\Delta$. Note that all derivable rules are
admissible, but not vice versa. A formula~$\fii$ is
\emph{admissibly saturated} \cite{ej:lukbas}, if every admissible rule
of the form~$\fii\ru\Delta$ is derivable. $\fii$ is \emph{exact} \cite{dj-ex}
if there exists a substitution~$\sigma$ such that
\[\fii\vdash\psi\iff\vdash\sigma(\psi)\]
for every formula~$\psi$.
$\fii$ is \emph{projective} \cite{ghilil} if it has a unifier~$\sigma$
(called a \emph{projective unifier}) such that
\[\fii\vdash p_i\eq\sigma(p_i)\]
for every~$p_i$. This implies that $\fii\vdash\psi\eq\sigma(\psi)$ for
every~$\psi$, and that $\sigma$ is an mgu of~$\fii$: if
$\tau\in U(\fii)$, we have $\tau\equiv\tau\circ\sigma$.

\pagebreak[2]
\begin{Fact}\label{fact:admsat}
Let $\fii$ be a formula.
\begin{enumerate}
\item\label{item:1} If\/ $\fii$ is projective, it is exact.
\item\label{item:2} If\/ $\fii$ is exact, it is admissibly saturated.
\end{enumerate}
\end{Fact}
\begin{Pf}
\ref{item:1}: On the one hand, $\sigma$ is a unifier of~$\fii$. On the
other hand, if $\vdash\sigma(\psi)$, then
$\fii\vdash\psi\eq\sigma(\psi)$ implies $\fii\vdash\psi$.

\ref{item:2}: If $\fii\adm\Delta$, then $\vdash\sigma(\psi)$ for
some~$\psi\in\Delta$ as $\sigma$ is a unifier of~$\fii$, hence
$\fii\vdash\psi$ by exactness.
\end{Pf}
A \emph{projective approximation} of~$\fii$ \cite{ghilil} is a finite
set~$\Pi$ of projective formulas such that
$\fii\adm\Pi$, and $\pi\vdash\fii$ for every
$\pi\in\Pi$. More generally, an
\emph{admissibly saturated approximation} \cite{ej:lukbas} is a set
with properties as
above, except that its elements are only required to be admissibly
saturated instead of projective. If $\Pi$ is an admissibly
saturated approximation of~$\ET\Gamma$, it is easy to see
(\cite[Obs.~3.7]{ej:lukbas}) that
\begin{equation}\label{eq:1}
\Gamma\adm\Delta\iff\forall\pi\in\Pi\,\exists\psi\in\Delta\,
  \pi\vdash\psi.
\end{equation}
If $\Pi$ is a projective approximation of~$\fii$, then the
set of projective unifiers of elements of~$\Pi$ is a finite complete set of
unifiers of~$\fii$. This does not hold for admissibly saturated
approximations in general.

The definition immediately implies that if $\Pi$ is any admissibly
saturated approximation of an admissibly saturated formula~$\fii$, then
there is a formula $\pi\in\Pi$ interderivable with~$\fii$ (i.e.,
$\fii\vdash\pi$ and~$\pi\vdash\fii$). In particular, if an admissibly
saturated formula has a projective approximation, it must be
projective itself, hence we have:
\begin{Fact}\label{fact:projapx}
The following are equivalent.
\begin{enumerate}
\item Every $\fii$ has a projective approximation.
\item Every $\fii$ has an admissibly saturated approximation, and
every admissibly saturated formula is projective.
\end{enumerate}
\end{Fact}
Projective formulas and approximations are the backbone of Ghilardi's
analysis \cite{ghil} of unification and admissibility in transitive
modal logics such as $\lgc{K4}$, $\lgc{S4}$, or~$\lgc{GL}$.
He shows that in these logics, every formula has a projective
approximation, which implies that unification is at most finitary, and
gives a description of admissibility by means of~\eqref{eq:1}.
By Facts \ref{fact:admsat} and~\ref{fact:projapx}, the same property also
implies that admissibly saturated, exact, and projective formulas coincide.

For an example exhibiting different behaviour, in \L ukasiewicz logic
every formula has an admissibly saturated approximation, and exact
formulas coincide with admissibly saturated formulas, but the logic
has nullary unification type, and some exact formulas are not
projective \cite{ej:lukbas,mar-spa,cab:exact}.

\section{Results}
As all of our results concern properties of the formula $p\to\Box p$,
our first task is to describe a complete set of unifiers of this
formula. Without further ado, this set will consist of the following
substitutions.
\begin{Def}\label{def:unif}
For any~$n\in\omega$, we introduce the substitutions
\begin{align*}
\sigma_n(p)&=\Box^{<n}p\land\Box^n\bot,\\
\sigma_\top(p)&=\top,
\end{align*}
where $\sigma_\alpha(q)=q$ for every variable~$q\ne p$ and
$\alpha\in\omega_+:=\omega\cup\{\top\}$.
\end{Def}
\begin{Lem}\label{lem:unify}
$\sigma_\alpha$ is a unifier of $p\to\Box p$ for
every~$\alpha\in\omega_+$.
\end{Lem}
\begin{Pf}
Using the principle $\fii\to\psi\vdash\Box^n\fii\to\Box^n\psi$, and
distributivity of~$\Box$ over~$\land$, we have
\begin{gather*}
\vdash\Box^{<n}p\land\Box^n\bot\to\Box^{\le n}p
  \to\Box\Box^{<n}p,\\
\vdash\Box^n\bot\to\Box^{n+1}\bot,
\end{gather*}
whence
$$\vdash\Box^{<n}p\land\Box^n\bot\to\Box\Box^{<n}p\land\Box\Box^n\bot
\to\Box(\Box^{<n}p\land\Box^n\bot).$$
Clearly, $\vdash\top\to\Box\top$.
\end{Pf}
We start with simple criteria for recognizing that a given unifier
of $p\to\Box p$ is below~$\sigma_\alpha$.
\begin{Lem}\label{lem:sign}
If $\sigma$ is a unifier of $p\to\Box p$, and~$n\in\omega$, the
following are equivalent:
\begin{enumerate}
\item\label{item:spsn} $\sigma\preceq\sigma_n$,
\item\label{item:sessn} $\sigma\equiv\sigma\circ\sigma_n$,
\item\label{item:boxn0} $\vdash\sigma(p)\to\Box^n\bot$.
\end{enumerate}
\end{Lem}
\begin{Pf}
\ref{item:sessn}${}\to{}$\ref{item:spsn} follows from the
definition of~$\preceq$.

\ref{item:spsn}${}\to{}$\ref{item:boxn0}: If
$\sigma\equiv\tau\circ\sigma_n$, then $\vdash\sigma_n(p)\to\Box^n\bot$
implies $\vdash\tau(\sigma_n(p))\to\tau(\Box^n\bot)$, i.e.,
$\vdash\sigma(p)\to\Box^n\bot$.

\ref{item:boxn0}${}\to{}$\ref{item:sessn}: Put~$\fii=\sigma(p)$.
Since $\sigma$ is a unifier of $p\to\Box p$, we have
$\vdash\fii\to\Box\fii$, hence $\vdash\fii\to\Box^{<n}\fii$ by
induction on~$n$. Since we also assume $\vdash\fii\to\Box^n\bot$, we have
$\vdash\sigma(p)\to\sigma(\sigma_n(p))$. The other implication is
trivial as~$\vdash\sigma_n(p)\to p$.
\end{Pf}
\begin{Def}
For any substitution~$\sigma$, let~$\sigma\res p$ be the
substitution~$\tau$ such that $\tau(p)=\sigma(p)$, and $\tau(q)=q$ for
every variable~$q\ne p$.
\end{Def}
\begin{Lem}\label{lem:sigtop}
If $\sigma$ is a substitution, the following are
equivalent:
\begin{enumerate}
\item\label{item:3} $\sigma\preceq\sigma_\top$,
\item\label{item:4} $\sigma\equiv\sigma\circ\sigma_\top$,
\item\label{item:5} $\sigma\res p\equiv\sigma_\top$,
\item\label{item:6} $\vdash\sigma(p)$.
\end{enumerate}
\end{Lem}
\begin{Pf}
\ref{item:4}${}\eq{}$\ref{item:5}${}\eq{}$\ref{item:6}: If $q\ne p$ is
a variable, we have $\sigma_\top(q)=(\sigma\res p)(q)=q$ and
$(\sigma\circ\sigma_\top)(q)=\sigma(q)$, hence the corresponding
equivalences in \ref{item:4} and~\ref{item:5} are trivially valid. For
$p$ itself, we have $\sigma_\top(p)=(\sigma\circ\sigma_\top)(p)=\top$,
hence \ref{item:4} and~\ref{item:5} both amount to
$\vdash\sigma(p)\eq\top$, which is the same as $\vdash\sigma(p)$.

\ref{item:4}${}\to{}$\ref{item:3} follows from the definition
of~$\preceq$. Conversely, if $\sigma\equiv\tau\circ\sigma_\top$, we
have $\tau(\sigma_\top(p))=\top$, thus $\vdash\sigma(p)$.
\end{Pf}
The crucial element in the description of $U(p\to\Box p)$ is to show
that one of the conditions in Lemma~\ref{lem:sign} or~\ref{lem:sigtop}
applies to every unifier. This amounts to a variant of the
rule of margins, as alluded to in the introduction. The basic idea is
similar to Williamson's proof \cite{will-alt} of the rule of margins for~$\lgc{KD}$: in order to
invalidate~$\fii\to\Box\fii$, we take two models satisfying $\fii$
and~$\neg\fii$, respectively, and join them by a path, while making
sure this does not mess up the valuation of~$\fii$ in the
end-points. Then $\fii$ has to switch to~$\neg\fii$ somewhere along
the path, at which point the formula $\fii\to\Box\fii$ will not hold.
\begin{Thm}\label{prop:rbm}
If\/ $\vdash\fii\to\Box\fii$, then $\vdash\fii$ or
$\vdash\fii\to\Box^n\bot$, where~$n=\md(\fii)$.
\end{Thm}
\begin{Pf}
Assume $\nvdash\fii$ and~$\nvdash\fii\to\Box^n\bot$. By
Fact~\ref{fact:tree}, the latter implies that there exists a finite
irreflexive intransitive tree $\p{F,R,\model}$ with root~$x_0$ such
that $F,x_0\model\fii\land\dia^n\top$. This means that there exists a
sequence $x_0\mr Rx_1\mr R\cdots\mr Rx_n$ of elements of~$F$, and as
$R$ is an intransitive tree, $x_n\notin R^{<n}(x_0)$.
Since~$\nvdash\fii$, there exists a model $\p{G,S,\model}$ and a
point~$x_{n+1}\in G$ such that~$G,x_{n+1}\nmodel\fii$. Let
$\p{H,T,\model}$ be the disjoint union of $F$ and~$G$, where we
additionally put~$x_n\mr Tx_{n+1}$. Since $F\res R^{\le n}(x_0)=H\res
T^{\le n}(x_0)$, we have $H,x_0\model\fii$ by Fact~\ref{fact:nbr}. On
the other hand, $H,x_{n+1}\nmodel\fii$, hence there exists~$i\le n$
such that $H,x_i\model\fii$ and~$H,x_{i+1}\nmodel\fii$. Then
$H,x_i\nmodel\fii\to\Box\fii$.
\end{Pf}
Ignoring the explicit dependence of~$n$ on~$\fii$, we can rephrase
Theorem~\ref{prop:rbm} by saying that the infinitary multiple-conclusion
rule
\begin{equation}\label{eq:2}
p\to\Box p\ru\{p\to\Box^n\bot:n\in\omega\}\cup\{p\}
\end{equation}
is admissible in~$\lgc K$. Let us mention that a similar proof
also shows that $\lgc K$ satisfies the following variant of
Williamson's alternative rule of disjunction: if
$n_0\ge\md(\fii_0)$, $n_1,\dots,n_k>\md(\fii_0)$, and
$\vdash\fii_0\lor\Box^{n_1}\fii_1\lor\dots\lor\Box^{n_k}\fii_k$, then
$\vdash\fii_0\lor\Box^{n_0}\bot$ or $\vdash\fii_i$ for some
$i=1,\dots,k$. We leave the details to the interested reader as we
have no further use for this property.
\begin{Cor}\label{prop:compl}
The substitutions $\{\sigma_\alpha:\alpha\in\omega_+\}$
form a complete set of unifiers of the formula $p\to\Box p$.
\end{Cor}
\begin{Pf}
By Lemmas \ref{lem:unify}, \ref{lem:sign}, and~\ref{lem:sigtop}, and Theorem~\ref{prop:rbm}.
\end{Pf}

\begin{Thm}\label{cor:nulunif}
Unification in~$\lgc K$ is nullary.
\end{Thm}
\begin{Pf}
Since $\vdash\sigma_n(p)\to\Box^{n+1}\bot$ and
$\nvdash\sigma_{n+1}(p)\to\Box^n\bot$,
Lemma~\ref{lem:sign} shows that $\sigma_n\prec\sigma_{n+1}$. Similarly,
$\nvdash\sigma_n(p)$ and $\nvdash\top\to\Box^n\bot$, hence
$\sigma_n$ and~$\sigma_\top$ are incomparable by
Lemmas \ref{lem:sign} and~\ref{lem:sigtop}. By Corollary~\ref{prop:compl}, every maximal
element of~$U(p\to\Box p)$ is equivalent to some~$\sigma_\alpha$, and
in view of $\sigma_n\prec\sigma_{n+1}$, we must have~$\alpha=\top$.
Thus, none of the unifiers~$\sigma_n$ is majorized by a maximal
element in~$U(p\to\Box p)$.
\end{Pf}
\begin{figure}
\centering
\includegraphics*{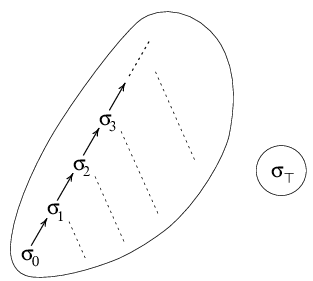}
\caption{Unifiers of $p\to\Box p$.}
\label{fig:unif}
\end{figure}
The preorder of unifiers of~$p\to\Box p$ is depicted in
Figure~\ref{fig:unif}. (We consider substitutions defined only
for the $p$~variable in the diagram, which is why there are no unifiers strictly
below $\sigma_\top$ or~$\sigma_0$.)

The basic description logic~$\alc$ \cite{bhs,baa-ghi} is a notational variant
of multimodal~$\lgc K$, with concept names corresponding to
propositional variables, and universal and existential restrictions
corresponding to boxes and diamonds, one pair for each role name. We
obtain immediately the following.
\begin{Cor}\label{cor:alc}
Unification in~$\alc$ is nullary, even for formulas with only one role
name and one concept name\footnote{That is, one concept variable and no concept constants. We employ no unification
problems with constants in this paper.}.
\noproof\end{Cor}

Now we turn to the (non)equivalence of exact and admissibly saturated
formulas. That $p\to\Box p$ is inexact follows easily from
Theorem~\ref{prop:rbm}:
\begin{Prop}\label{prop:exact}
The formula $p\to\Box p$ is not exact, and a fortiori not projective.
\end{Prop}
\begin{Pf}
Assume for contradiction that $\sigma$ is a substitution such that
\[p\to\Box p\vdash\psi\iff\vdash\sigma(\psi)\]
for every~$\psi$. In particular, $\sigma$ is a unifier of $p\to\Box
p$, hence $\vdash\sigma(p)$ or $\vdash\sigma(p)\to\Box^n\bot$ for
some~$n$ by Theorem~\ref{prop:rbm}. However, $p\to\Box p\nvdash p$ and
$p\to\Box p\nvdash p\to\Box^n\bot$, a contradiction.
\end{Pf}
We remark that $\sigma_n$ and~$\sigma_\top$ are projective unifiers
of the formulas $p\to\Box p\land\Box^n\bot$ and~$p$, respectively.

We complement Proposition~\ref{prop:exact} by showing that $p\to\Box p$ is
admissibly saturated. We mention another pathological property of
$p\to\Box p$ which will arise from the proof. Intuitively, it is not
so surprising that a formula~$\fii$ with an infinite cofinal chain of
unifiers like~$\sigma_n$ (or more generally, a formula whose preorder of unifiers is directed, even if it has no
maximal element) can be admissibly saturated, as the unifiers
high enough in the chain eventually become ``indistinguishable'' when
applied to any particular formula~$\psi$. However, if a formula has
two incomparable maximal unifiers, say~$\sigma,\sigma'$, we would
expect it \emph{not} to be admissibly saturated: presumably, we can
find formulas~$\psi,\psi'$ unified by $\sigma$ and~$\sigma'$,
respectively, but not vice versa. Then $\fii\adm\psi,\psi'$, but not
$\fii\adm\psi$ or~$\fii\adm\psi'$. By the same intuition, we would expect
that a formula like $p\to\Box p$, whose set of unifiers consists of
two incomparable parts (a chain and a maximal unifier, in our case),
is not admissibly saturated either.

What happens here is that when we apply the unifiers~$\sigma_n$ to a
particular formula, they not only become ``indistinguishable'' from
each other for $n$~large enough, but they also ``cover'' the
unifier~$\sigma_\top$, despite that it is not comparable to any
element of the chain. Returning to our weak rule of margins, one can
imagine that the margins of error about the approximate
falsities~$\Box^n\bot$ gradually blend into the margin about the
truth~$\top$ as $n$~goes to infinity.
\begin{Prop}\label{prop:admsat}
The formula $p\to\Box p$ is admissibly saturated.
\end{Prop}
\begin{Pf}
Assume $p\to\Box p\adm\Delta$, and pick
$n>\max\{\md(\psi):\psi\in\Delta\}$. Since $\sigma_n$ unifies
$p\to\Box p$, there exists~$\psi\in\Delta$ such
that~$\vdash\sigma_n(\psi)$. We claim
\[p\to\Box p\vdash\psi.\]
If not, there exists a Kripke model~$\p{F,R,\model}$ such that
$F\model p\to\Box p$ and~$F,x_0\nmodel\psi$ for some~$x_0\in F$.
First, we unravel~$F$ to a tree (cf.\ \cite[Prop.~2.15]{brv}, \cite[Thm.~3.18]{cha-zax}): let $\p{G,S,\model}$ be the model
where $G$ consists of sequences $\p{x_0,\dots,x_m}$ such
that~$m\in\omega$, $x_i\in F$, $x_i\mr Rx_{i+1}$; we put
$\p{x_0,\dots,x_m}\mr S\p{x_0,\dots,x_m,x_{m+1}}$; and
$G,\p{x_0,\dots,x_m}\model p_j$ iff $F,x_m\model p_j$ for each
variable~$p_j$. The mapping $f\colon G\to F$ given
by~$f(\p{x_0,\dots,x_m})=x_m$ is a p-morphism, hence it preserves the
valuation of formulas by Fact~\ref{fact:pmor}. In particular, $G\model
p\to\Box p$ and~$G,\p{x_0}\nmodel\psi$.

Let $H$ be the submodel of~$G$ consisting of sequences
$\p{x_0,\dots,x_m}$ where~$m<n$. We still have $H\model p\to\Box p$:
if $\vec x=\p{x_0,\dots,x_m}$ with $m<n-1$, then $G\res S^{\le1}(\vec x)=H\res
S^{\le1}(\vec x)$, hence $H,\vec x\model p\to\Box p$ by
Fact~\ref{fact:nbr}; on the other hand, if $m=n-1$, then $H,\vec x\model\Box\bot$, and a fortiori $H,\vec x\model p\to\Box p$.
It follows that $H\model p\to\Box^{<n}p$, and
moreover $H\model\Box^n\bot$, hence $H\model p\eq\sigma_n(p)$. However, $G\Sset H\Sset G\res S^{\le\md(\psi)}(\p{x_0})$, hence
$H,\p{x_0}\nmodel\psi$ by Fact~\ref{fact:nbr}. These
properties together imply $H,\p{x_0}\nmodel\sigma_n(\psi)$,
contradicting~$\vdash\sigma_n(\psi)$.
\end{Pf}
We remark that unlike Theorem~\ref{prop:rbm}, we could not directly take a
finite irreflexive intransitive tree for~$F$ in the proof above,
because $\lgc K$ is not finitely strongly complete with respect to
such frames. (Every finite irreflexive tree is converse well-founded,
and therefore validates L\"ob's rule~$\Box p\to p\ru p$, which is
admissible but not derivable in~$\lgc K$.)
\begin{Cor}\label{cor:projapx}
The formula $p\to\Box p$ has no projective approximation.
\end{Cor}
\begin{Pf}
In view of Propositions \ref{prop:exact} and~\ref{prop:admsat}, this follows from the discussion leading to Fact~\ref{fact:projapx}.
\end{Pf}

\section{Conclusion}
We have provided examples confirming that
unification and admissibility in the basic modal logic~$\lgc K$
involves peculiar phenomena not encountered in the familiar case of
transitive modal logics with frame extension properties:
the fact that $\lgc K$ has the worst possible unification type,
even for very simple formulas in one variable like $p\to\Box p$, is a
problem by itself; as we have seen, this formula is also a
counterexample to other structural properties vital for the kind of
analysis of admissibility and unification that has been applied in the
transitive case, namely it is neither projective nor exact despite
being admissibly saturated, it has no projective approximation, and it
is admissibly saturated even though its preorder of unifiers is not
directed (it consists of two disjoint connected components).

The major remaining problem in this area is whether admissibility or
unifiability in~$\lgc K$ is decidable. Our results might be
seen as hinting towards the possibility that these tasks are undecidable.
(The results of Wolter and Zakharyaschev~\cite{wolt-zakh:undec} also
point in this direction.) For example, \eqref{eq:2} means that a rule of the form $\Gamma,p\to\Box p\ru\psi$ is admissible iff
$\Gamma,p\ru\psi$ and $\Gamma,p\to\Box^n\bot\ru\psi$ are admissible for every~$n\in\omega$. Note that $p\to\Box^n\bot$
holds in a model iff the submodel generated by points satisfying~$p$ is well-founded of finite depth at most~$n$; one
can imagine that the discrete nature of such models could be used to encode finite computation or some kind of finite
combinatorial structures. Since $n$ can be arbitrarily large irrespective of the size of $\Gamma$ or~$\psi$, this
might lead to an undecidable problem.

On the other hand, should admissibility in~$\lgc K$ be decidable after all,
our results show that proving this will require methods more powerful and more delicate than
what we are used to from the transitive case, as current techniques
are not ready to cope with obstacles exhibited by the behaviour
of~$p\to\Box p$.

\section{Funding}
This work was supported by grant IAA100190902 of GA AV \v CR,
project 1M0545 of M\v SMT \v CR, a grant from the John Templeton
Foundation, and RVO: 67985840.

\section{Acknowledgement}
I would like to thank the anonymous referee for useful suggestions.

%\bibliographystyle{mybib}
%\bibliography{kunif}
\providecommand{\bysame}{\leavevmode\hbox to5em{\hrulefill}\thinspace}

\end{document}